\documentclass[twocolumn,showpacs,preprintnumbers,amsmath,amssymb,
superscriptaddress]{revtex4}

\usepackage{graphicx}
\usepackage{dcolumn}
\usepackage{bm}

\begin{document}

\preprint{}

\title{Magnetic and Superconducting Properties of Underdoped Ba(Fe$_{1-x}$Co$_x$)$_2$As$_2$ Measured with Muon Spin Relaxation/Rotation}


\author{T.~J.~Williams}
\author{A.~A.~Aczel}
 \affiliation{Quantum Condensed Matter Division,
   Neutron Sciences Directorate,
   Oak Ridge National Lab,
   Oak Ridge, TN, 37831, USA}

\author{S.~L.~Bud'ko}
\author{P.~C.~Canfield}
 \affiliation{Department of Physics and Astronomy and Ames Laboratory, 
   Iowa State University, 
   Ames, Iowa 50011, USA}

\author{J.~P.~Carlo}
 \affiliation{Department of Physics,
   Columbia University,
   538 W. 120th St., New York, NY, 10027, USA}
 \affiliation{Department of Physics,
   Villanova University,
   Villanova, PA, 19085, USA}
   
\author{T.~Goko}     
\author{Y.~J.~Uemura}
 \affiliation{Department of Physics,
   Columbia University,
   538 W. 120th St., New York, NY, 10027, USA}

\author{G.~M.~Luke}
 \affiliation{Department of Physics and Astronomy,
   McMaster University, 1280 Main St. W.,
   Hamilton, ON, L8S 4M1, Canada}

\date{\today}

\begin{abstract}
We report muon spin relaxation/rotation ($\mu$SR) measurements of single 
crystal Ba(Fe$_{1-x}$Co$_x$)$_2$As$_2$ with $x$=0.038 and 0.047. Zero field 
(ZF)-$\mu$SR and Transverse field (TF)-$\mu$SR measurements of these 
underdoped samples find the presence of magnetism and superconductivity.  
We find internal fields along the $\hat{c}$-axis whose magnitude decreases 
with increasing doping.  We find evidence for a low-temperature volume 
fraction that is only weakly magnetic, where that volume fraction increases 
with increasing Co doping of the sample.  TF-$\mu$SR measurements show 
slight changes in the spectra that indicate magnetic inhomogeneities due to 
the loss of Fe moments in the system, the effect of which is larger in the 
higher Co doping.  We discuss the existence of superconductivity in these 
samples in close proximity to strong magnetic order.
\end{abstract}


\maketitle

\section{Introduction}
Of the various families of iron pnictide superconductors, the
so-called 122 family has been extensively studied due to their high
T$_C$'s and the ability to grow relatively large single crystals, which 
includes BaFe$_2$As$_2$, SrFe$_2$As$_2$ and CaFe$_2$As$_2$. Unlike the case 
of the cuprates, superconductivity in these materials is apparently quite 
robust against in-plane disorder, brought about by electron-doping for Fe 
atoms either by Co, Ni or other transition metals.  The transition 
temperatures remain fairly high for these substitutions, with T$_C$ = 22K 
for Ba(Fe$_{0.926}$Co$_{0.074}$)$_2$As$_2$~\cite{Sefat_08c,Ni_08},
20.5K for Ba(Fe$_{0.952}$Ni$_{0.048}$)$_2$As$_2$~\cite{Canfield_09}, 23K for
Ba(Fe$_{0.9}$Pt$_{0.1}$)$_2$As$_2$~\cite{Saha_10}, 14K for 
Ba(Fe$_{0.961}$Rh$_{0.039}$)$_2$As$_2$~\cite{Ni_09}, 19.5K for 
Sr(Fe$_{0.8}$Co$_{0.2}$)$_2$As$_2$~\cite{Kim_09}, 9.5K for 
Sr(Fe$_{0.925}$Ni$_{0.075}$)$_2$As$_2$~\cite{Saha_09} and 12K for 
Ca(Fe$_{0.972}$Co$_{0.028}$)$_2$As$_2$~\cite{Ran_12}.

Like the cuprate superconductors, these materials exist in close proximity 
to magnetism.  However, the nature of the competition or coexistence of 
magnetism and superconductivity in the pnictides has been debated, 
particularly since there may be different behaviours in different 
families~\cite{Lumsden_10}.  This continues with the nature of the gap 
symmetry which also may vary across different families~\cite{Reid_12} and 
where the temperature-dependence of the superfluid density raises the 
possibility of multi-band superconductivity~\cite{Williams_09,Terashima_09}.

\section{Experimental}
Muon spin rotation ($\mu$SR) is a powerful local microscopic tool for 
characterizing the magnetic properties of materials in superconducting or 
other states. A thorough description of the application of $\mu$SR to 
studies of superconductivity can be found elsewhere~\cite{Sonier_07}. Each 
implanted muon spin precesses around the local magnetic field until the muon 
decays into a positron, which is preferentially ejected along the direction 
of the muon spin at the time of decay (as well as two neutrinos which are 
not detected). Due to the large gyromagnetic ratio of the muon, local fields 
as small as 0.1G can be detected.

Single crystals of Ba(Fe$_{1-x}$Co$_{x})_2$As$_2$ with x=0.038 and 
0.047 were grown at Ames from self flux as described in detail 
elsewhere~\cite{Ni_08}.  The crystals, each of roughly 1cm$^2$ area, were 
mounted in a helium gas flow cryostat on the M15 and M20 surface muon
beamlines at TRIUMF, using a low background arrangement such that only 
positrons originating from muons landing in the specimens were collected 
in the experimental spectra.

\section{Magnetism and Coexistence}
\begin{figure}[thb]
\begin{center}
\includegraphics[angle=0,width=\columnwidth]{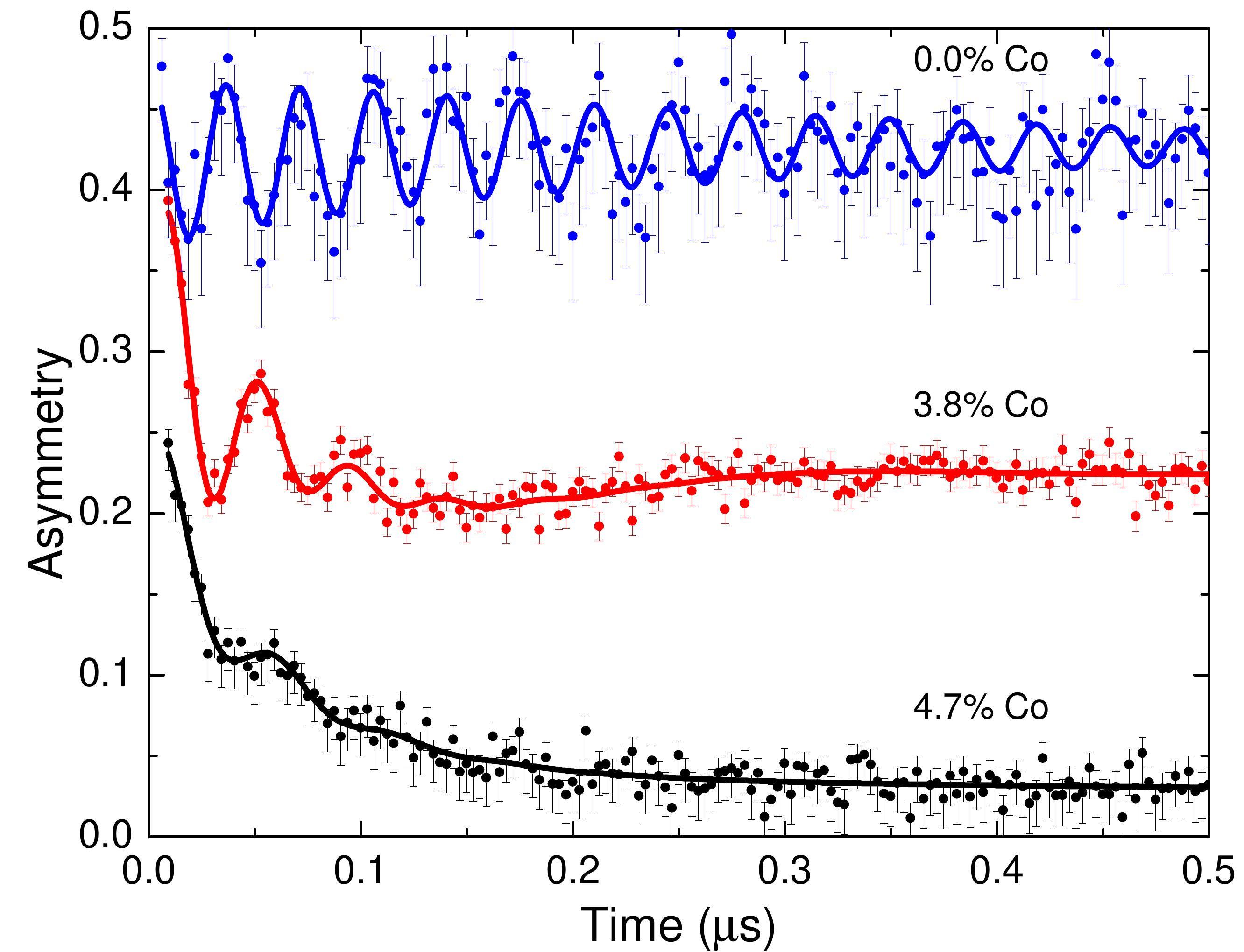}
\caption{\label{rawdata}
(Color online)
The zero-field (ZF)-$\mu$SR spectra taken at T=1.65K for the parent (x=0, 
blue), x=0.038 (red) and x=0.047 (black) samples.  The x=0.038 data is 
vertically offset by +0.15 and the x=0 data is vertically offset by +0.3 for 
clarity.  The solid lines are the fits to the data, as described in the 
text.  Here, we see clear precession in the parent material, but the 
precession in the doped compounds is strongly damped, indicating 
significantly disordered magnetism.}
\end{center}
\end{figure}

Using Zero Field (ZF)-$\mu$SR, magnetic fields at the muon site can be 
measured through precession of the muon spins in the local field.  The 
presence of magnetic order in the samples is manifested as an oscillation in 
the ZF-$\mu$SR spectra, as was the case in the samples measured here.  
Characteristic spectra for the samples with x=0.038, 0.047 and the parent 
material, BaFe$_2$As$_2$, at T=1.65K are shown in Fig.~\ref{rawdata}, with 
the spectra vertically offset for clarity.  Here and elsewhere in this work, 
the data for the parent compound was originally reported in 
Ref.~\cite{Aczel_08}.  Motivated by previous studies of the underdoped and 
parent compounds in which two muon sites have been observed 
\cite{Marsik_10,Aczel_08}, as well as DFT calculations supporting two 
electrostatic minima in the crystal structure~\cite{DeRenzi_12}, the 
ZF-$\mu$SR data was fit to a fitting function:
\begin{equation}A(t)=
A_1\cos(\omega_1t)*e^{-\lambda_1t}+A_2\cos(\omega_2t)*e^{-\lambda_2t}+A_ee^{-\lambda_et}
\label{eq1}
\end{equation}

This is a model containing two precessing muon signals and a relaxing, 
non-precessing signal.  Furthermore, a single damped exponential did not fit 
the data well.  In Fig.~\ref{rawdata}, we can see the clear precession 
indicative of well-ordered magnetism that is present in the parent material 
is not long-lived in the doped samples.  The highly damped precession 
indicates a broad distribution of local fields at the muon sites, indicating 
a substantial degree of disorder in the magnetic ordering.

\begin{figure}[thb]
\begin{center}
\includegraphics[angle=0,width=\columnwidth]{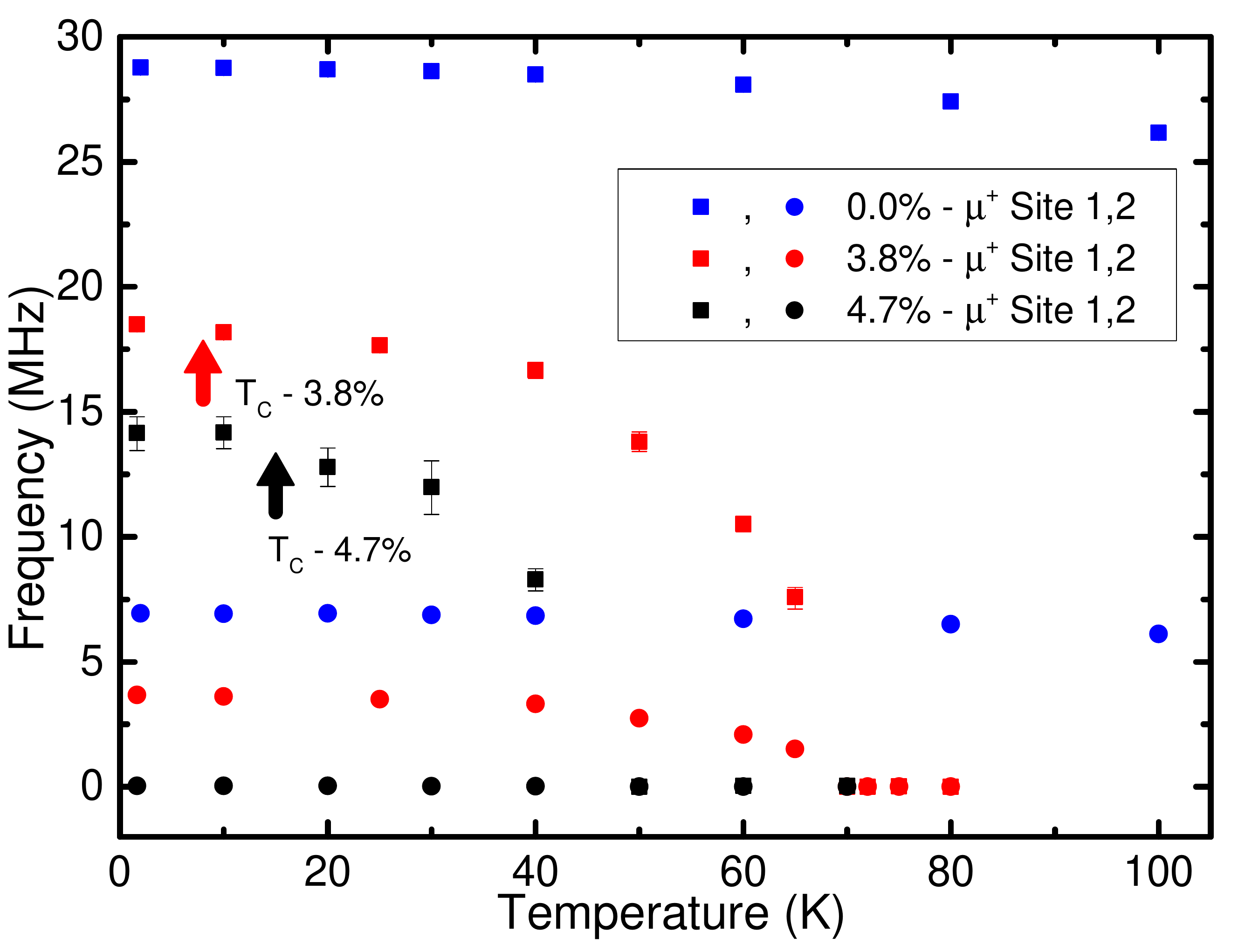}
\caption{\label{zf_frq}
(Color online)
(a) ZF-$\mu$SR spectra for Ba(Fe$_{1-x}$Co$_x$)$_2$As$_2$ with x=0.038 
(T$_{N}$=71K and T$_C$=8K) and x=0.047 (T$_{N}$=45K and T$_C$=15K).  
There is a clear onset at T$_{N}$, where the precession frequency becomes 
non-zero.  There is no significant change below T$_C$.}
\end{center}
\end{figure}

The precession frequencies, $\omega_1$ and $\omega_2$, extracted from this 
model are shown as a function of temperature in Fig.~\ref{zf_frq}. The onset 
of the magnetic ordering was estimated to be 71(3)K for 
Ba(Fe$_{0.962}$Co$_{0.038}$)As$_2$ and 45(2)K for 
Ba(Fe$_{0.953}$Co$_{0.047}$)As$_2$, in excellent agreement with transport 
and scattering measurements~\cite{Ni_08,Pratt_09}.

In these measurements, we find that the magnitudes of the two frequencies 
scale with one another, which is evidence for two inequivalent muon sites 
with different local fields.  This scaling would not be expected in the case 
of muons landing in different regions of the sample with distinct 
environments, such as multiphase samples.  The scaling dependence was found 
during the initial analysis and so the frequencies and asymmetries of the two 
signals were constrained to be a constant multiple of one another.  These 
ratios were fit to a temperature-independent value using the whole 
temperature range below T$_N$.  The frequency ratio was found to decrease 
with doping, being $0.24(1)$ for the parent material, $0.20(4)$ at 3.8\% Co 
doping and $2.3(4)\times 10^{-3}$ at 4.7\%. The decrease in the ratio may 
reflect the growing Co concentration, which distorts the 
magnetically-ordered lattice and the electrostatic potential at the muon 
site~\cite{DeRenzi_12}.  At the 4.7\% doping, the extremely small value of 
the ratio of high- and low-field sites may represent the nearly complete loss 
of long-range order, so that all sites appear to have different local fields, 
consistent with the data from Fig.~\ref{rawdata}, where the oscillations 
disappear within the first 0.1-0.2$\mu$s.  The frequencies observed in our 
measurements are lowest in the 4.7\% sample, while both are lower than for 
the parent compound\cite{Aczel_08}.  This indicates that the average magnetic 
field at both muon sites decreases with increasing doping.  This could be due 
to the replacement of Fe moments with Co atoms, which reduces the size of the 
internal field.  The replacement of Fe moments with Co atoms may also cause 
disorder in the local field, leading to a high muon relaxation rate in the 
ZF-$\mu$SR measurements.  As seen in Fig.~\ref{rawdata}, the relaxation rates 
of the oscillating components are highest in the 4.7\% sample and slightly 
lower in the 3.8\% sample, while the oscillations in the parent material 
persist to much longer times.  This is in agreement with other work on the 
underdoped Ba(Fe$_{1-x}$Co$_x$)$_2$As$_2$ compounds, where the precession 
signals observed are fairly short-lived, indicating that the magnetic 
ordering in these samples are most likely strongly disordered or 
incommensurate\cite{Marsik_10}.  Incommensurate magnetic order has been 
observed by neutron scattering in dopings above $x$=0.056~\cite{Pratt_11}. 

When ZF-$\mu$SR spectra are measured with the initial muon polarization in 
the direction of the local field, $\vec{P}_{\mu} \parallel \vec{H}_{loc}$, 
the spectra do not relax to zero asymmetry in the ordered region.  This 
behaviour was observed when the direction of the initial muon spin was 
parallel to the crystallographic $\hat{c}$ direction.  This was verified by 
measuring ZF-$\mu$SR spectra in an orientation in which the muon spin is 
perpendicular to the $\hat{c}$-axis, $\vec{P}_{\mu} \perp \vec{H}_{loc}$, 
where the spectra do relax to zero at long times.  The fits of the ZF-$\mu$SR 
spectra give internal fields at the high-field muon site to be approximately 
0.13T in x=0.038 and 0.11T in x=0.047.  This is consistent with other studies 
of Ba(Fe$_{1-x}$Co$_x$)$_2$As$_2$ that have found strong internal fields 
($>$0.15T) along the $\hat{c}$-axis in various dopings, from the parent 
compound up to the loss of magnetic order\cite{Marsik_10}.

We also measured these samples in a transverse field (TF) of 4mT to probe the 
paramagnetic volume fraction.  In a weakly magnetic or paramagnetic region 
the muon ensemble sees a local field equal to the applied field, $H_{app}$, 
so it can be fit to a simple exponentially-damped cosine:

\begin{equation}A(t)=
A\cos(\omega t)*e^{-\lambda t}
\label{eq2}
\end{equation}

\noindent
where $\omega = \gamma_{\mu} H_{app}$.	

\begin{figure}[htb]
\begin{center}
\includegraphics[angle=0,width=\columnwidth]{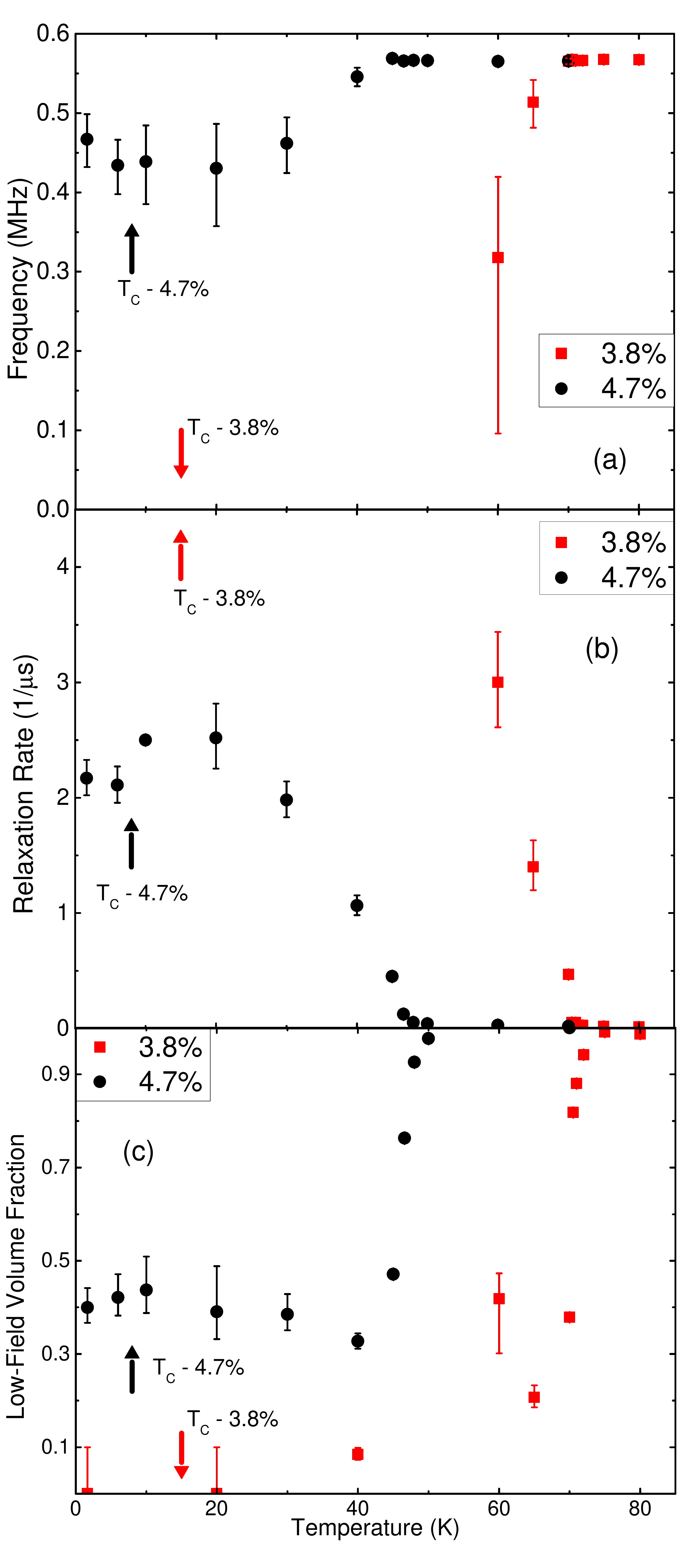}
\caption{\label{wtf_frq_rlx}
(Color online) TF-$\mu$SR measurements of x=0.038 and 0.047 in TF=4mT.
(a) The frequency shows a drop below the magnetic ordering temperature, but 
slowly increases again towards T=0.  This may be indicative of field-induced 
ordering. 
(b) The relaxation rate increases sharply below T$_N$, saturating well before 
T=0.  This indicates that the local field is fairly inhomogeneous in these 
samples, which causes dephasing of the muon spin polarization in the ordered 
part of the phase diagram. 
(c) We see a 100\% paramagnetic signal above T$_{N}$, which drops sharply at 
the transition.  Below the ordering temperature, there is a residual 
paramagnetic signal, about 40\% for x=0.047 and less than 10\% for x=0.038.  
Due to the small residually-precessing volume fraction present in the 3.8\% 
sample, the points below 50K are omitted from (a) and (b).}
\end{center}
\end{figure}

Below a magnetic ordering transition, the local field becomes a vector sum 
of the ordered and external fields, so the muons no longer precess at the 
frequency given by the external field.  The amplitude of the measured signal, 
$A$, of the precessing signal decreases, though any para- or weakly-magnetic 
region will still precess at this frequency.  This allows TF-$\mu$SR to be 
used for measurements of the magnetic volume fractions in 
magnetically-ordered materials.  In these samples, we find that the volume 
fraction of the magnetically-ordered region increases sharply below T$_{N}$ 
(as shown in Fig~\ref{wtf_frq_rlx}(c)), saturating within 15K of the 
ordering temperature.  We see that in the 4.7\% sample, there is a residual 
signal that still precesses down to the lowest temperatures measured, but 
has a relaxation that increases below T$_{N}$ indicating that it is 
weakly magnetic.  This low-field region is less than 10\% in the x=0.038 
sample, while it is approximately 40\% for x=0.047.  This is an indication 
that the fraction of the sample that sees weak magnetism increases with 
increasing Co concentration.

Fig.~\ref{wtf_frq_rlx}(a) and (b) show the precession frequencies, $\omega$,  
and relaxation rates, $\lambda$, present in the TF-$\mu$SR spectra.  The 
relaxation rate measures the rate of the depolarization of the muon spins, 
which may be caused by inhomogeneities in the local magnetic field among 
other factors.  We see in Fig.~\ref{wtf_frq_rlx}(a) that the frequency shows 
a decrease below the magnetic ordering transition.  This can be attributed to 
the antiferromagnetic order that emerges and causes inhomogeneities in the 
local field; this is significant evidence for disordered magnetism at these 
dopings.  This inhomogeneity in the local field is also seen in the 
TF-$\mu$SR relaxation rates (Fig.~\ref{wtf_frq_rlx}(b)) that show a sharp 
increase at the magnetic ordering transition.  The relaxation rates saturate, 
and remain relatively constant down to $T=0$.  When this is taken together 
with the ZF and with the TF-$\mu$SR spectra in different orientations, we 
find a magnetic structure that displays a large internal field oriented along 
the $\hat{c}$-axis.  As was seen with $\mu$SR measurements of the parent 
material\cite{Aczel_08}, these samples see magnetism everywhere, but in the 
doped samples the magnetism is highly disordered.  This is in agreement with 
NMR\cite{Laplace_09,Ning_09} and M\"{o}ssbauer\cite{Bonville_10} measurements 
that find disordered, possible incommensurate magnetic order in the 
underdoped Ba(Fe$_{1-x}$Co$_x$)$_2$As$_2$ compounds.  

This is somewhat in contrast to neutron scattering measurements that see 
well-defined magnetic scattering, which continues to be sharp below 
T$_C$\cite{Pratt_10}.  Neutron scattering also sees incommensurate 
magnetic order at higher dopings, for x=0.056 to x=0.060\cite{Pratt_11}.  
This incommensuration depends on composition, though the magnetic Bragg peaks 
remain well-defined.  The fact that neutron scattering sees a well-ordered 
magnetic structure may be due to fluctuations in the local magnetism on 
timescales slower than that detectable by neutron scattering, but observable 
by $\mu$SR.  This is in agreement with other techniques that operate on 
similar timescales, such as M\"{o}ssbauer ($\sim 10^{-7}$s) and NMR ($\sim 
10^{-5}$s).  Incommensuration has been observed at dopings near the loss of 
antiferromagnetic order with a very small splitting~\cite{Pratt_11}.  In 
lower doped samples, such as those in this study, it is possible that the 
incommensuration is too small to be observed by neutron scattering but 
results in disorder in the local field, consistent with what has been 
observed in the Ni-doped system~\cite{Arguello_13}.  Another possible 
explanation is that the magnetic moments are a superposition of an ordered 
and a disordered component.  While neutron scattering is sensitive only to 
the ordered component, muons would see both components, resulting in a 
largely disordered local field.  This is consistent with neutron measurements 
of Ca(Fe,Co)$_2$As$_2$, which find a decrease in the size of the Fe ordered 
moment with increasing Co-doping~\cite{Lester_09,Prokes_11}.  In the latter 
two cases, the disorder would increase with increasing Co-doping, as is seen 
in both ZF- and TF-$\mu$SR relaxation rates.

Below the superconducting transition, we see only small changes in the 
TF-$\mu$SR parameters.  We see a slight decrease in the relaxation rate in 
the 4.7\% sample below T$_C$=15K, while we do not have enough data points 
around and below T$_C$=8K in the 3.8\% sample to see any changes in 
behaviour.  By using a higher transverse field of 0.02T, we find more direct 
evidence of superconductivity.  This was measured at T=1.65K for both 
field-cooled (FC) and zero field-cooled (ZFC) orientations.  Compared to the 
FC data, the ZFC spectrum displayed a strong increase in the relaxation, 
indicative of flux pinning, clearly indicating that we have superconductivity 
in these samples.  Superconductivity may exist in the low-field regions since 
this region exhibits a growing volume fraction with increasing doping and by 
x=0.061 there is no static magnetism present\cite{Williams_10}.  The entire 
sample displays magnetism but some regions have small local fields, so there 
may be phase separation on the nanoscale with superconductivity existing in 
or near the magnetic regions of the sample.  This is in agreement with 
scanning tunnelling spectroscopy (STS) results, that see no phase separation 
on length scales larger than tens of nanometers\cite{Massee_10}.

\section{Conclusions}
We have measured single crystals of Ba(Fe$_{1-x}$Co$_x$)$_2$As$_2$ that show 
both magnetism and superconductivity using Zero-Field and Transverse-Field 
muon spin relaxation/rotation.  These measurements demonstrate the clear 
onset of a local magnetic field below T$_N$, largely along the 
$\hat{c}$-axis, the magnitude of which decreases with increasing Co doping.  

TF-$\mu$SR measurements suggest some inhomogeneity in the local field, where 
the disorder is greater in the 4.7\% doping as evidenced by the lower 
precession frequency in the TF-$\mu$SR signal at low T and a higher 
relaxation in the ZF-$\mu$SR spectra.  There is a residual volume fraction 
that displays low-field magnetism, which increases with Co doping from less 
than 10\% in BaFe($_{0.962}$Co$_{0.038}$)$_2$As$_2$ to 40\% in 
BaFe($_{0.953}$Co$_{0.047}$)$_2$As$_2$.  A possible source of this disorder 
may be the loss of Fe moments, which results in lower internal fields in the 
higher doped sample.  These measurements are consistent with other techniques 
that suggest an incommensuration of the magnetic order that increases with 
Co-doping.  This is reflected in the ZF-$\mu$SR measurements, that show 
reduced precession frequencies and increased relaxation rates at higher 
dopings, suggesting lower internal fields and increased magnetic disorder.

Measurements in a larger transverse field of 0.02T display flux pinning and 
superconductivity, which may display nanoscale phase separation existing in 
the low-field regions of the sample.  However, there is no detectable change 
in the local magnetic field or magnetic volume fractions below T$_C$.  
This suggests that the entire sample sees some magnetic order, indicating 
that the superconductivity exists in or near regions of large local magnetic 
order.

\

The authors acknowledge helpful earlier work on this project done by N. Ni.  
We also appreciate the hospitality of the TRIUMF Centre for Molecular and 
Materials Science where the majority of these experiments were performed. 
Research at McMaster University is supported by NSERC and CIFAR. Work at 
Columbia was supported by NSF-DMR-0502706 and NSF-DMR-0806846.
Work by P.C.C. and S.L.B. at Ames Laboratory was supported by the Department 
of Energy, Basic Energy Sciences under Contract No.
DE-AC02-07CH11358. 








\end{document}